# Nature, Science, and PNAS – Disciplinary profiles and impact

Staša Milojević

*Center for Complex Networks and Systems Research, Luddy School of Informatics, Computing, and Engineering, Indiana University, Bloomington*

## Abstract

*Nature*, *Science*, and *PNAS* are the three most prestigious general-science journals, and *Nature* and *Science* are among the most influential journals overall, based on the journal Impact Factor (IF). In this paper we perform automatic classification of ~50,000 articles in these journals (published in the period 2005-2015) into 14 broad areas, to explore disciplinary profiles and to determine their field-specific IFs. We find that in all three journals the articles from Bioscience, Astronomy, and Geosciences are over-represented, with other areas being under-represented, some of them severely. Discipline-specific IFs in these journals vary greatly, for example, between 18 and 46 for *Nature*. We find that the areas that have the highest disciplinary IFs are not the ones that contribute the most articles. We also find that publishing articles in these three journals brings prestige for articles in all areas, but at different levels, the least being for Astronomy. Comparing field-specific IFs of *Nature*, *Science* and *PNAS* to other top journals in six largest areas (Bioscience, Medicine, Geosciences, Physics, Astronomy, and Chemistry) these three journals are always among the top seven journals, with *Nature* being at the very top for all fields except in Medicine.

## INTRODUCTION

The publication of the first two scientific journals, *Journal des Sçavans* and *Philosophical Transactions*, in France and England, respectively, almost simultaneously in 1665, marked the beginning of fundamental changes in the way that burgeoning scientific research was communicated. The first journals were not only quite different from the ones we are familiar with today, in numerous ways, but they also differed from each other in terms of "contents and their intentions" (Meadows, 1998, p. 6). The proliferation of journals, which started in the late 19[th] century and continues to this day, has led to the establishment of different roles for this genre and, more importantly, has been instrumental in shaping scientific norms (Zuckerman & Merton, 1971).

Contemporary science has witnessed an exponential growth in both the number of papers (Fortunato et al., 2018; Price, 1961, 1963) and in the number of journals (Meadows, 1998; Price, 1961, 1974), leading to close to 100,000 scientific journals covering all the scientific and scholarly fields across the world (Ioannidis, 2006). The exponential growth in scientific literature has been accompanied by increased specialization (Meadows, 1998; Price, 1961). As the knowledge became more specialized, we saw the proliferation of specialized journals meeting the needs of the communities around these specialized topics (Ziman, 1969). This has led to a wide variety of journals when it comes to the level of their specialization (Glänzel, Schubert, & Czerwon, 1999). However, a relatively small number of journals still cover a wide range of topics (general science journals). Several of those journals are considered particularly prestigious and they tend to publish what the scientific or broader community sees are important breakthroughs or new ideas (Ackerson & Chapman, 2003; Palmer, 1996).

Journals are not only the vehicles to communicate the latest findings or to serve as archives of cumulative past claims – a "relatively robust archive of humanity's scientific knowledge" (Csiszar, 2018, p. 1). Publication of original research in a scientific journal has also been used to: (a) establish priority of findings (Johns, 1998); (b) signal belonging to certain intellectual communities; (c) establish relative standing within scientific communities; and (d) differentiate professional scientists from "laypeople" (Csiszar, 2018). It is therefore not surprising that journals have played a pivotal role in academic rewards and professional recognition of individuals. Namely, prestige of a journal has often been used, implicitly if not explicitly, as an assessment of the quality of research (De Rijcke, Wouters, Rushforth, Franssen, & Hammarfelt, 2016; Ravetz, 1971).

In the complex landscape of science communication, career pathways, rewards, and research funding highly prestigious journals, *Nature* and *Science*, and to a lesser extent *PNAS*, play an important role. There is an intricate relationship between audience, authors, and perceived journal prestige that has the potential to lead to positive feedback loops, giving disproportionate advantage to certain journals. Meadows (1998) has identified "the regard in which a journal is held by its research community" (p. 164) and "the audience reached by the journal" (p. 164) as two basic factors driving submission decisions for authors across all disciplines. And, these publishing practices "seem to be the nerve center where issues of reward, responsibility, and status merged on an everyday basis" (Csiszar, 2018, p. 11).

Studies have shown that disproportionate number of highly cited papers tend to concentrate in the top general science journals, especially the *Nature* and *Science* (Ioannidis, 2006). At the same time, we know that these journals do not cover all areas of science equally (Ackerson & Chapman, 2003; Ding, Ahlgren, Yang, & Yue, 2018; Kaneiwa et al., 1988) – they tend to publish papers from the fields that have the highest number of average citations per paper (Ioannidis, 2006). These journals may also exhibit a "chaperone effect" (Sekara et al., 2018), making it difficult for the authors who have not published in these journals before to do so. In any case, there is a consensus that these journals represent good vehicles to disseminate work to broader audiences (Ackerson & Chapman, 2003) and potentially increase its impact. In the era when science is experiencing an exponential growth in number of publications leading to "attention deficit", journals are used for helping researchers locate relevant information leading to disproportional relying on general science and high-impact field journals (De Rijcke et al., 2016; Rushforth & De Rijcke, 2015).

Today, the prestige of a journal is almost exclusively measured and discussed based on the Journal impact factor (IF). The IF is essentially an average number of citations recently published in some journal. IF has been developed by Eugene Garfield (1972) as a measure designed to select journals to be included in the newly founded citation databases (Reedijk & Moed, 2008). Selection of specific journals was important because at the time the computing resources were limited and expensive. The measure soon left the realm of information retrieval, and moved to science evaluation circles. However, despite widely spread usage of an IF as a measure of journal success, a large body of research has argued that IF does not capture all the complexity of evaluating the impact of journals (Bar-Ilan, 2012; Bornmann, Werner, Gasparyan, & Kitas, 2012; Haustein, 2012; R. Rousseau, 2002; Thelwall, 2012). The most contested application of IF has been using it as a proxy for evaluating authors, via the IF of journals in which their papers have appeared (Archambault & Larivière, 2009; DORA, 2012). This practice was justified by the need for "immediacy" in evaluation, when individual work is so recent that it hasn't had time to accrue citations, but it is often used even when citation data on individual articles are extensive. The principal deficiency of the IF is twofold: citation distributions are right skewed with power laws (Seglen, 1992, 1997) and citation distributions are also very broad and overlapping even for journals with very different IFs (Larivière et al., 2016; Milojević, Radicchi, & Bar-Ilan, 2017; Stringer, Sales-Pardo, & Amaral, 2008). These two characteristics make IF a poor predictor of the number of citations that an individual paper will receive, that is, its actual impact.

Despite these known limitations of the IF, there are some indicators that authors rely on the impact of the journal when deciding where to submit their work (Garfield, 2006; S. Rousseau & Rousseau, 2012), with many authors believing that having their work published in a higher-impact venue will lead to more rewards, via increased visibility and potentially more citation (Calcagno et al., 2012) and ultimately higher values of performance measures such as an h-index. Since some of the general science journals tend to have very high impact factors, the authors often opt for them rather than the more specialized journals (Verma, 2015). These pressures are occasionally leading to the "cascading" of the submissions, with the authors starting with the highest-impact journal and getting down the hierarchy until their paper is accepted (Gordon, 1984), adding burden to the system.

While we do know that general science journals such as *Nature* and *Science* have among the highest IFs of all journals (Fang, 2015) (IF of around 30), for the high IF to potentially translate into increased impact one should ask whether the papers published in those high impact journals fair equally well regardless of the *discipline* of the article, or do different disciplines have different impacts? More specifically, are high impact factors driven by high rates of citations in specific disciplines? Because we deal only with an IF of an entire journal, the answer to the question what individual "impact factors" different disciplines in these journals have is not straightforward. Related to that, there is a question, of how much "benefit", in terms of citation, is there in publishing in general-science high profile journals, rather than in top disciplinary journals? And, if there is, are the benefits universal, or discipline dependent?

A principal obstacle to providing the answers to these question lies in the fact that the large bibliographic databases only contain disciplinary classification at the level of journals, not articles. Performing a classification of tens of thousands of articles required for a statistically robust analysis is not trivial, and requires a use of automated methods. Specifically, in the Web of Science, one of the major bibliographic databases from the publisher of IF, the journals such as *Nature*, *Science* and *PNAS* are classified in a category called "multidisciplinary". This name can be confusing, because it is not that the individual articles in those journals are necessarily multidisciplinary themselves, rather these journals carry disciplinary articles from a large number of disciplines (Hicks & Katz, 1996; Katz & Hicks, 1995; Waltman & van Eck, 2012). .

Classification of scientific literature is of utmost importance for both descriptive and evaluative science studies. A number of researchers suggested methods to reclassify individual articles published in the general science journals (i.e., journals in the "Multidisciplinary sciences" subject category in Web of Science (WoS)). Many of the proposed solutions are based on the references of the articles (e.g., Glänzel & Schubert, 2003; Glänzel, Schubert, & Czerwon, 1999; Glänzel, Schubert, Schoepflin, & Czerwon, 1999; López-Illescas, Noyons, Visser, De Moya-Anegón, & Moed, 2009). A more recent solution to this problem utilized both citing and cited publications as basis for reclassification (Ding et al., 2018). Recently, we have developed a method to reclassify the entire WoS database at a level of articles, regardless of the subject category assigned to the journal in which the article was published. Our reclassification is into the ~240 existing WoS categories, excluding the non-specific ones labeled as multidisciplinary. Furthermore, for high-level studies such as the present one, we also reclassify the articles into 14 broad areas (disciplines). In this study we will explore disciplinary composition of *Nature*, *Science*, and *PNAS* across these 14 areas and derive individual "impact factors" in these disciplines and compare them to some high-impact disciplinary journals.

## DATA AND METHODS

In this paper we use reference-based classification which, similar to a number of previous classifications, employs an article's references to infer its topic. To perform the classification we initially use only the references that were published in journals that have a single subject category which is not "multidisciplinary" (i.e., it is not published in multidisciplinary or general disciplinary journals). We refer to such items as classifier references or *classifiers*. The tallying of the subject categories of classifiers allows us to determine the unique WoS subject category of items that were published in general subject journals. After the first reclassification we repeat the process, but now using the newly assigned subject categories of references. Reclassification is based on the full Web of Science (WoS) Core Collection database containing items published from 1900 through the end of 2017. The database contains a total of 69 million items (bibliographic entries), of which 55 million have at least one reference recorded in the database. We perform the classification on (and using) all document types. The edition of WoS used in this work uses 252 subject categories. For higher-level classification, we place each of 252 subject categories into 14 broad areas. Names of broad areas are taken from NSF WebCASPAR Broad Field (Javitz et al., 2010). Details and classification algorithm are described in (Milojević, revision under review). Evaluation of the classification showed a very high level of reliability – errors do not exceed ~1% for classification in broad areas.

In this paper we introduce a new measure, which we call the *prestige of a journal* which is defined as

$$P = \frac{IF_{journal}}{IF_{WoS}}$$

or, the ratio of an IF-like measure for a single journal (or its disciplinary component) and an IF-like measure for all WoS articles in a given discipline. This measure facilitates a normalization for field dependent IFs in a way similar to the one undertaken for normalization of citation distributions (Radicchi, Fortunato, & Castellano, 2008).

For this study, we focus on articles published in a ten-year period (2005-2015) in three major general science journals *Nature* (9,261 articles classified into individual fields), *Science* (8,844 articles), and *PNAS* (39,169 articles) (Table 1). These articles received 571,371 (*Nature*), 451,013 (*Science*) and 706,945 (*PNAS*) citations over the period 2007-2016. Parts of the analyses required comparisons to articles published in different fields from all other sources (mostly journals). There were 14,076,076 such items and 59,284,270 citations they have received, over the same time periods. For the analysis we select items classified in WoS as articles. We do not include review papers that tend to skew the IFs and introduce an asymmetry in the analysis because some journals publish exclusively reviews, some published them to some extent and many not at all.

Table 1. Summary of the data used in this study. IF(JCR) stands for the Impact Factor of journals reported in the Journal Citation Reports and IFe stands for the Impact Factor *estimated* from the Web of Science data. Both values are the means for 2007-2016.

|  | Articles | Articles classified | Citations received | <IF> (JCR) | <IFe> |
|---|---|---|---|---|---|
| Nature | 9327 | 9261 | 571,371 | 36.8 | 35.3 |
| Science | 8873 | 8844 | 451,013 | 31.5 | 29.1 |
| PNAS | 39,173 | 39,169 | 706,945 | 9.6 | 10.2 |
| All sources | 15,317,691 | 14,076,076 | 59,284,270 | - | 2.2 |

The purpose of this study is not to rank the journals, but to better understand the disciplinary profiles of the top science journals and how each discipline contributes to the impact of the journal as a whole. Earlier studies have shown (e.g., Milojević et al., 2017) that these three journals have quite similar citation distributions. This similarity is especially strong between *Science* and *Nature*. Furthermore, the pair-wise comparison of citation capacity of journals ("citation success index") based on citation distributions has tight relation to the IF ratio of the journals being compared (Milojević et al., 2017). So, for simplicity and familiarity of the audiences, we will use an IF-like measure in this study.

The IF is a very simple metric. Namely, the IF of a journal in year y equals the number of citations received in y to all documents published in that journal in the preceding two years ($y - 2$ and $y - 1$), divided by the number of "citable documents" covered by the citation database (Moed & van Leeuwen, 1996). Given that we are interested in the disciplinary contributions to the given IFs we cannot be using IFs reported in the Journal Citation Reports, but have calculated our own (IFe in Table 1). The correlation between the official IFs and IFe is very strong (Figure 1) and the small discrepancies are well-known and reported in the literature (e.g., Bar-Ilan, 2010). Specifically, we have applied a small correction factor of 1.04 (MRB) to account for the fact that the official measure includes citations to all items in the numerator, but only certain document types (articles, reviews) in the denominator. Since we are not including reviews in our analysis, our IFs will in some cases be smaller than the official ones because review articles tend to be cited more highly on average than research articles.

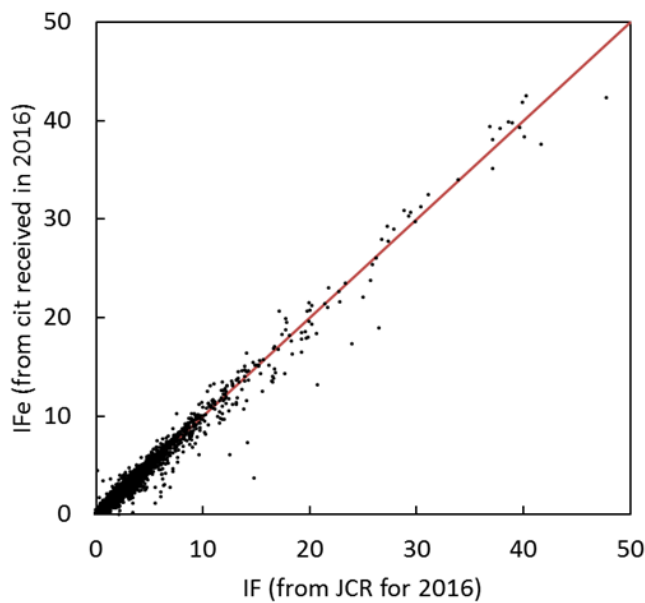

Figure 1. Correlation between the official Impact Factors reported in the 2016 Journal Citation Reports and Impact Factors estimated (IFe) in the Web of Science using citations received in 2016. IFe data here include review articles, but the ones we use in the analyses do not.

## RESULTS

***Relative contribution of different disciplines to high impact general science journals***

As stated in the introduction, it is well known that the general science journals do not have equal coverage of all the scientific fields. For example, in a study that examined articles published in *Nature* and *Science* over the period 1981-1983, Kaneiwa et al. (1988) found *Nature* and *Science* to exhibit similar disciplinary coverage, with around 50% of articles covering Medical science and Bioscience research. In a more recent study, Ding et al. (2018) used WoS data to analyze disciplinary profiles of *Nature, Science,* and *PNAS* in two periods (2004-2006 and 2014-2016). This study found Bioscience to be dominant in all three journals, followed by Medicine. In addition, it found *Science* and *Nature* to be more inclusive in their coverage as compared to *PNAS*, which had a higher concentration and Bioscience and Medicine articles than the other two.

In this study, we looked at the relative contribution of 14 broad areas of science to the research being published in *Nature*, *Science*, and *PNAS*. As can be seen in Figure 2, which shows the contributions of different disciplines in the three journals (average percentages for the three journals), most of the articles published in all three journals come from Bioscience (53%), followed by Medicine (17%), Geosciences, (9%) and Physics (9%). Only two more areas rise above 1% (Astronomy and Chemistry). All other areas, including Mathematics, Computer science, Humanities, Agriculture and Professional fields have very low coverage.

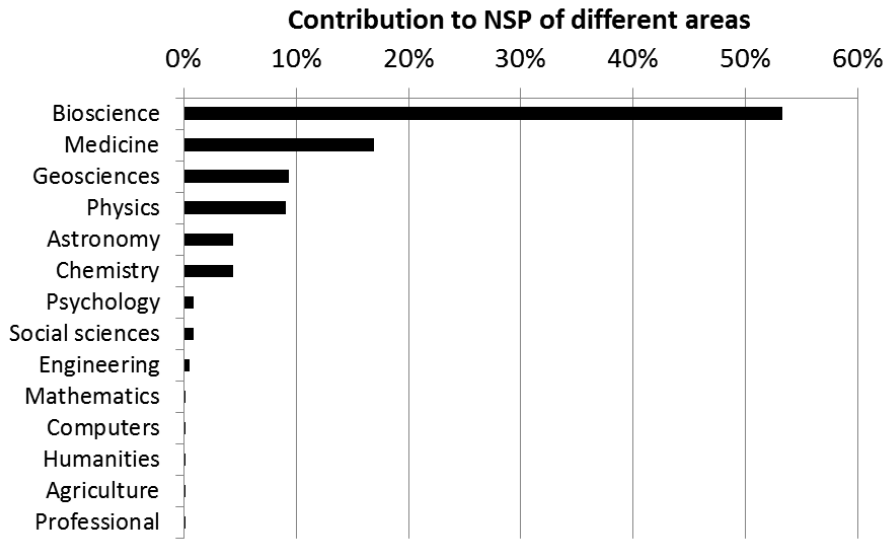

Figure 2. Disciplinary composition of articles published in *Nature*, *Science*, and *PNAS* (NSP) belonging to 14 broad areas. Contributions in each of three journals was averaged.

The interpretation of this quite uneven distribution as a preference to publish certain areas cannot be made based on these data alone. We would not expect the fractions to be uniform because different areas do not necessarily produce the same number of articles in general, and the fields that are most represented may simply be the largest. To test this we took into account the size of each of the broad areas based on all articles in the WoS. Interestingly, the disciplinary composition of the papers published in the top three general science journals is not representative of their relative contribution in scientific literature (Figure 3). Three areas are over-represented, with Bioscience being the most over-represented by a factor of 4, followed by Astronomy by a factor of 2 and to a smaller extent Geosciences (30%). Two other areas that show up as the top contributors are both under-represented, with Physics being more under-represented (70%) than Medicine (40%). Unsurprisingly, the areas that suffer from most underrepresentation are those for which we have seen very little presence among the reported research: Professional fields, Humanities, Computer science and Engineering, among others, which are underrepresented by factors of between 10 and 50.

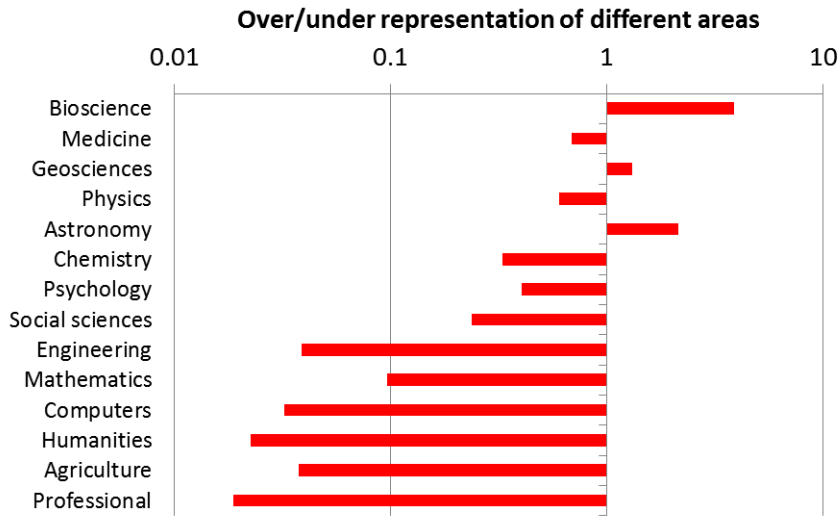

Figure 3. Overrepresentation (>1) or underrepresentation (<1) of individual broad areas in *Nature*, *Science*, and *PNAS* (combined) with respect to all articles in Web of Science database. The scale is logarithmic. Areas are sorted by their contributions in the three journals (Figure 2).

While the disciplinary composition of *Nature*, *Science*, and *PNAS* is generally similar, there are some interesting differences between them (Figure 4). While all three journals have very strong presence of Bioscience articles, *PNAS* has the highest fraction (62%) and *Science* the lowest of the three (45%). *Nature* and *Science* are very similar in terms of their coverage of Medicine, while *PNAS* has significantly higher coverage of this area (almost two times) than the other two. *Science* and *Nature* are also very similar in their coverage of Geosciences and Physics, whereas *PNAS* has significantly lower coverage of these areas (~3 times). This is especially true for Astronomy, where *PNAS* has close to no coverage at all.

We now come to the areas where even *Science* and *Nature* depart in their coverage. *Science* has the highest coverage of Chemistry articles, with *Nature* and *PNAS* being similar at close to third of *Science*'s coverage of the area. Looking at the filed with small overall contribution in any journal, we see that *Science* and *PNAS* are very similar in their (limited) coverage of Psychology, which is significantly higher (6 times) than in *Nature*. *Science* has the highest coverage of Engineering, with much lower, but mutually similar coverage of this area by *Nature* and *PNAS*. Finally, *PNAS* has the highest coverage of Mathematics of the three, with *Nature* not covering this area at all and *Science* only sporadically (3 articles in 10 years). The remaining four areas have such small contributions that no meaningful comparison is possible.

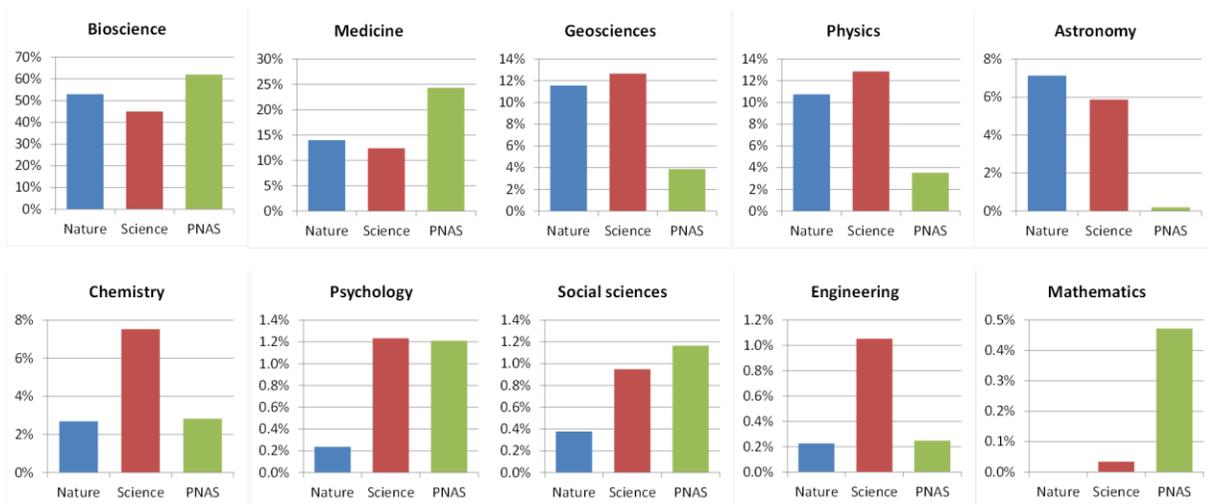

Figure 4. Shares of articles from ten different research areas in *Nature*, *Science*, and *PNAS*.

*Impact-factor-like measure by field*

Like other citation-based measures, IF is discipline/field dependent. It is affected both by the number of papers and by the number of references and it is known that both of these vary significantly by field. So, given the variety of areas present in a general science journal, a relevant question is what an IF-like measure for each individual broad area of research would be, and how it would compare to the overall IF of that journal. Furthermore, it would be informative to see whether broad areas exhibit similar tendencies across three journals.

Figure 5 shows the IF-like values (we'll refer to them as IF, for short) of 9 areas with non-negligible contribution in *Nature*. The order of rows follows the share of articles, from highest to lowest. First, we see that there is a significant range in individual disciplines: from IF = 18 for Geosciences and Psychology, to IF = 46 for Medicine and Chemistry – a span of a factor of 2.5. Disciplinary IFs tend to cluster in two groups: low, around 20 and high, around 40. There is no correlation between the IF and the level of contribution of that field in the journal. For example, Chemistry contributes only 3% of articles, whereas Medicine, contributes 14% of articles, but both have the similarly high IF. Two other fields have disciplinary IFs above the overall IF: Physics, which contributes 11% of articles and Bioscience, which is the largest contributor to articles with 53%.

Moving onto *Science* (Figure 6), we again have a wide range of disciplinary IFs, spanning a factor of 2.4. The split into which areas have similarly higher and lower IFs is the same as for *Nature*, except that Engineering is closer to the higher IF group. Interestingly, Chemistry, which in *Science* has much higher presence than in *Nature* (Figure 4), also has the highest IF, followed by Physics and Medicine (Figure 6). Otherwise, as in the case of *Nature*, there is no correlation between IF and contribution. *Nature* and *Science* have similar IFs, and *Nature*, which has a slightly higher IF, is outperforming *Science* in all broad areas except in Engineering and Geoscience.

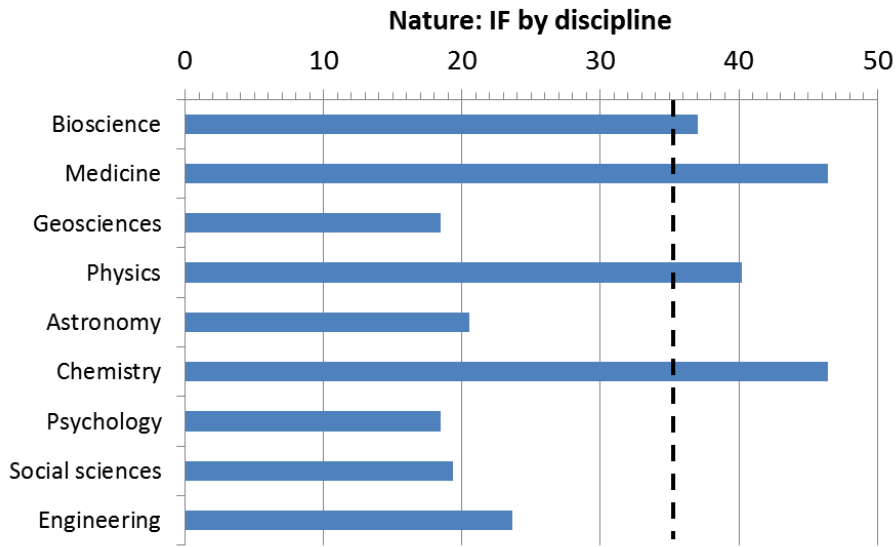

Figure 5. Impact factor-like measure (IF, for short) for 9 largest broad areas in *Nature*. Vertical dotted line shows the IF for the whole journal.

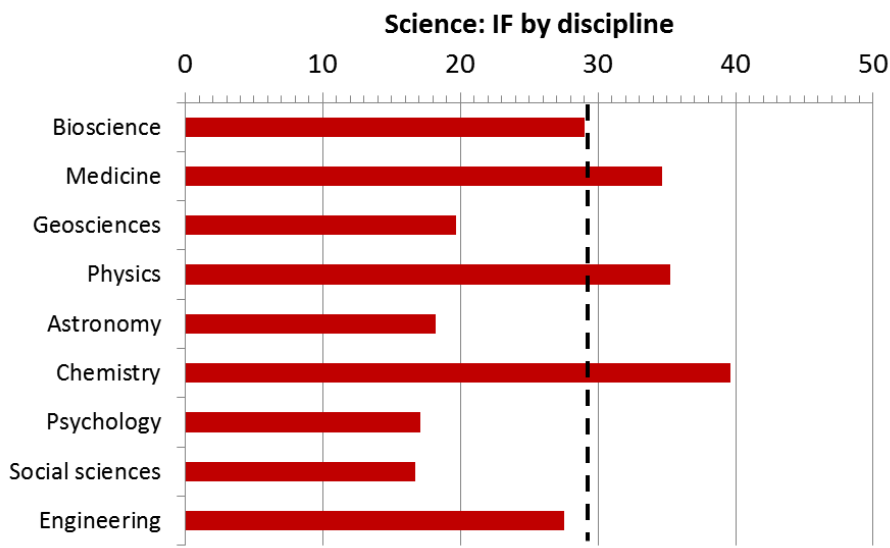

Figure 6. Impact factor-like measure (IF, for short) for 9 largest broad areas in *Science*. Vertical dotted line shows the IF for the whole journal.

*PNAS* has ~3 times lower IF than the other two journals and it also differs in terms of relative IFs between the disciplines (Figure 7). Furthermore, in *PNAS* we can make meaningful estimates for the IFs of the remaining 5 areas (Mathematics, Computer science, Humanities, Agriculture and Professional fields). Amongst the 9 areas in common with *Nature* and *Science*, there is a much smaller range of IFs (factor of 1.7). The relative range expands greatly when including the five lesser areas in terms of contribution – Agriculture has 6x higher IF than the Humanities. The broad areas that have the highest

IF-like measure in *PNAS* are Agriculture and Professional fields, two fields with only a very small contribution – 20 articles each over the ten years. Other broad areas whose IFs are above the IF for the whole journal (though not by much) are: Medicine, Geosciences, and Chemistry.

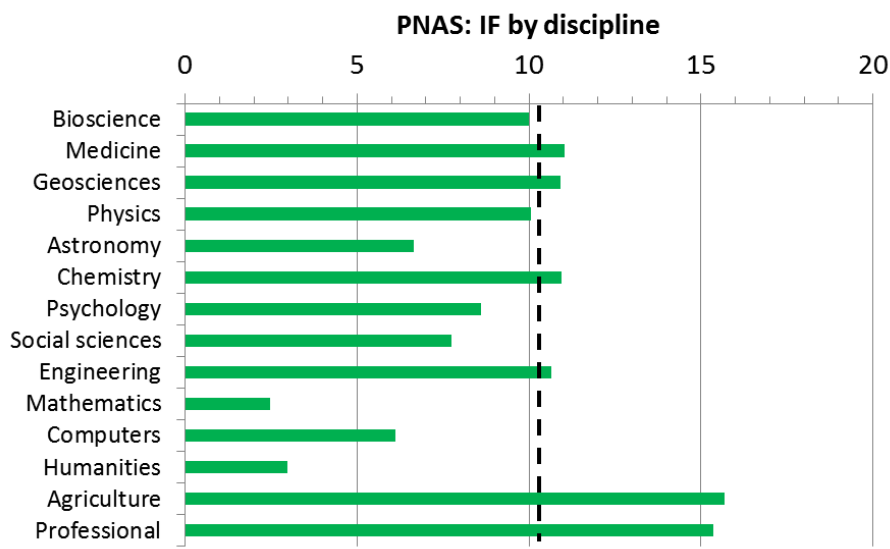

Figure 7. Impact factor-like measure (IF, for short) for all 14 broad areas in *PNAS*. Vertical dotted line shows the IF for the whole journal.

While the results presented above are informative, and we see that *Nature* and *Science* follow similar trends in disciplinary IFs, they lack the larger context. How do these IF-like measures for broad areas compare to the IF-like measure of the entire broad area, represented by all articles in the WoS, what we call the measure of prestige? Prestige effectively normalizes for disciplinary citation differences allowing us to better assess how much "better" does some area do with respect to that area in general. Interestingly, the prestige varies both across the fields and across the three journals. Publishing an article in *Nature* is actually most prestigious for Social sciences papers (P = 19, i.e. 19 times the typical IF) (Figure 8). This is followed by Medicine (P = 17 times), and Physics (P = 16 times). Astronomy papers published in *Nature* have the smallest level of prestige (P=5), the reason for which is that in Astronomy the typical IF of all articles is already relatively high (IF = 4.1) and since we have seen that Astronomy is overrepresented in *Nature*, the papers that get published there fail to be that much more exclusive.

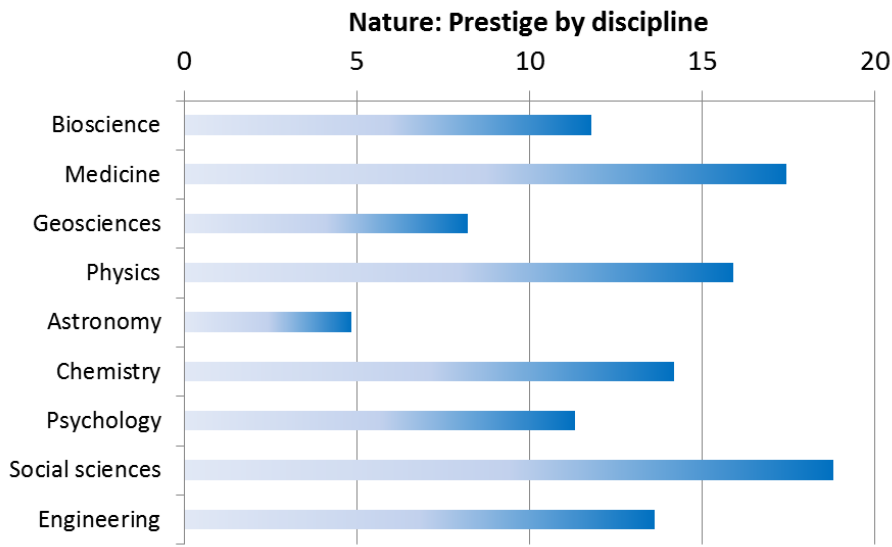

Figure 8. Prestige of articles for 9 broad areas published in *Nature*. Prestige is determined by dividing the Impact Factor for the *Nature* articles belonging to a broad area by the Impact Factor of all articles in that area indexed in the Web of Science.

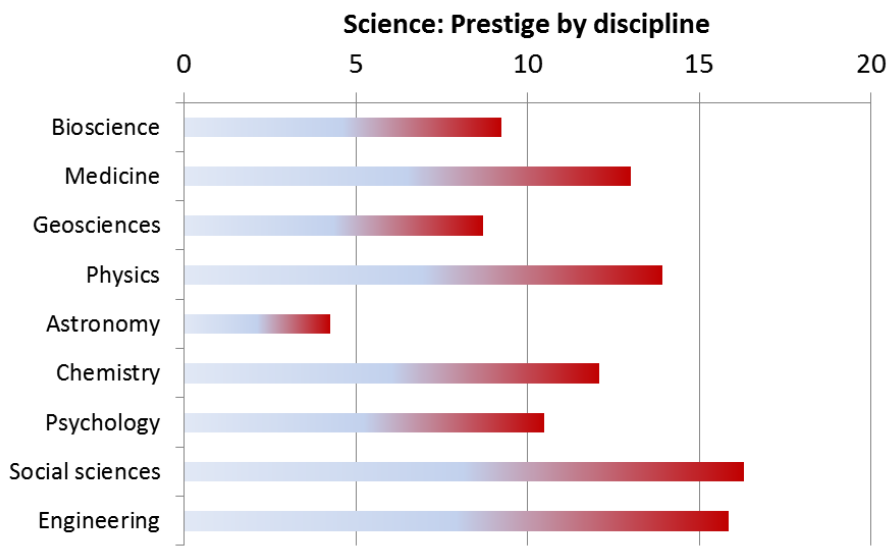

Figure 9. Prestige of articles for 9 broad areas published in *Science*.

Turning to the prestige of different disciplines in *Science* (Figure 9), articles there tend to again follow similar patterns to those of *Nature* since they have similar IFs by discipline. Still there are differences, Social sciences and Engineering articles enjoy the highest level of prestige (P = 16), followed by Engineering (16 times), followed by Physics, Medicine and Chemistry with P= 14, 13 and 12, respectively. Again, Astronomy articles would have the smallest prestige coefficient of only 4.

Similar to the results looking at IF by broad area, *PNAS* is different from *Nature* and *Science* in terms of the areas whose articles have the highest prestige. First, the overall P numbers are smaller because the IFs are smaller in PNAS. Top areas in terms of prestige are Professional fields (P = 14), Agriculture (P =12) and Humanities (P = 11) (Figure 10). High levels of prestige in these areas that rival the values for some areas in *Nature* and *Science* may be because *Nature* and *Science* do not publish articles in these areas, so the articles that would otherwise be *Nature* or *Science* "worthy" get published in *PNAS*. On the opposite end of the spectrum, similar to *Nature* and *Science*, Astronomy articles published in *PNAS* have the smallest level of prestige (P = 1.6) or just 60% above the IF for the whole area.

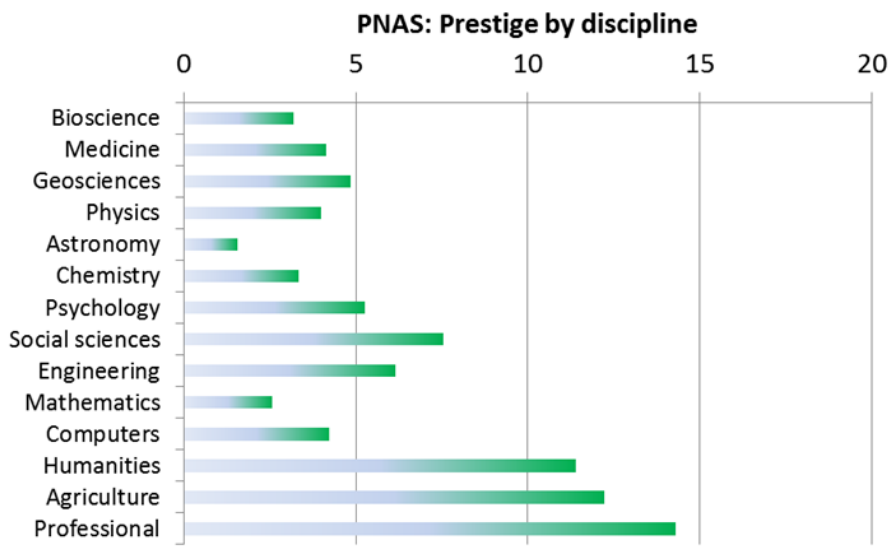

Figure 10. Prestige of articles for 14 broad areas published in *PNAS*.

***Impact in general science journals as compared to impact in top disciplinary journals***

The above analyses have shown that there are clear citation advantages for all the broad areas to publish articles in the top general science journals compared to a typical disciplinary venue. However, one would like to know to what degree does this hold when one focuses on the most prominent specialized/disciplinary journals. To test this we have compared IF of different broad areas in *Nature, Science*, *PNAS* and top disciplinary journals (Figure 11). Our selection for inclusion focused on journals that are relatively broad in that area while not publishing mostly reviews. We focus on six areas that are the greatest contributors to *Nature*, *Science* and *PNAS*. Altogether there are only few journals that rival *Nature* and *Science*, and some dozen ones that rival *PNAS*. In the overall landscape of journals this is still a small number.

In all but one instance *Nature* is the top venue. Only in Medicine, that place is taken by *New England Journal of Medicine*, with *Nature* taking the second place. In six fields (Geosciences, Physics, Astronomy, and Chemistry) *Science* is ranked second. It is ranked third in Bioscience and fourth in Medicine. *PNAS* is ranked third in Geosciences and Astronomy. It is ranked 5 in Physics, 6 in Chemistry, and 7 in Bioscience and Medicine.

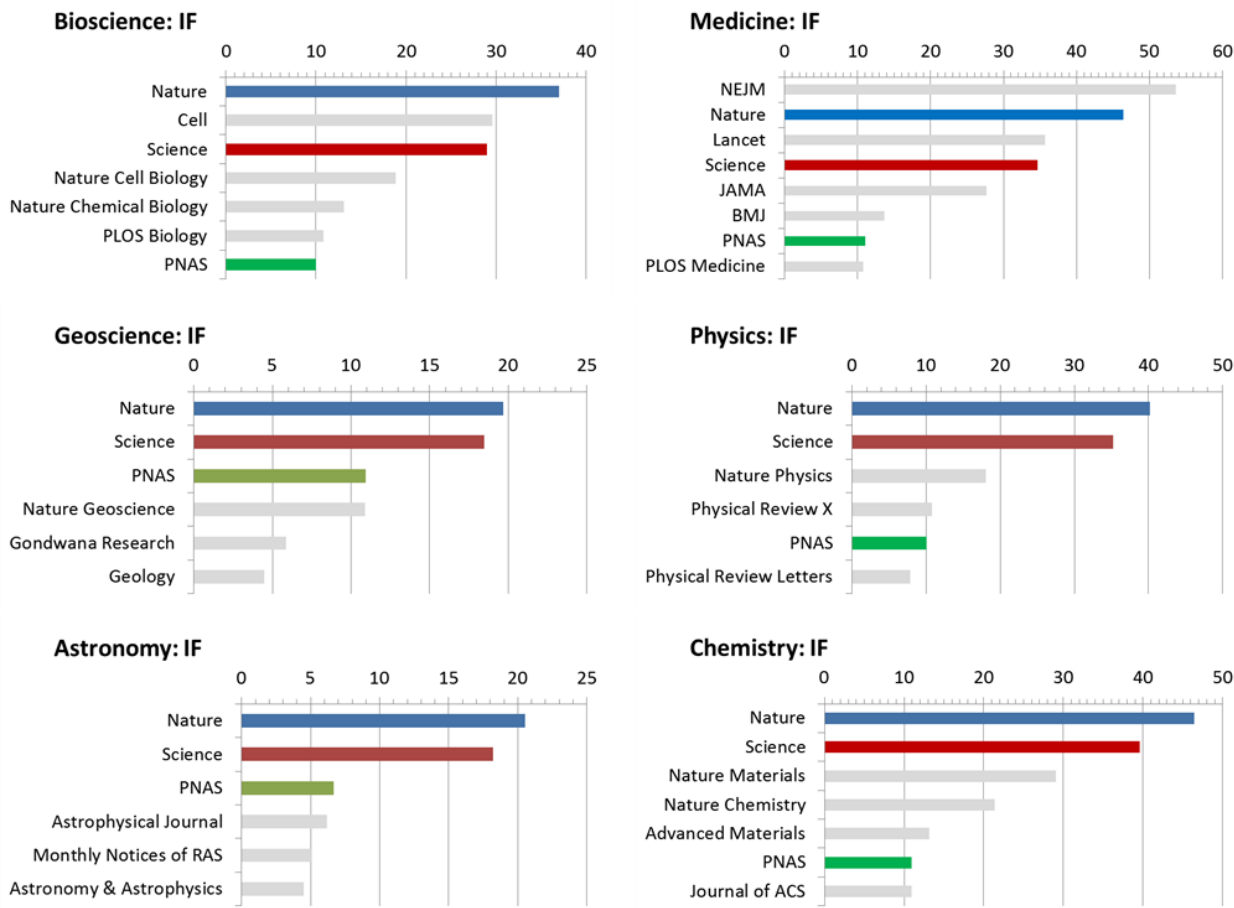

Figure 11. Ranking of journals based on the Impact Factor, with *Science*, *Nature*, and *PNAS* Impact Factors calculated only for the specified broad research area.

## DISCUSSION & CONCLUSIONS

The study has investigated disciplinary composition of articles published in top three general science journals (*Nature*, *Science* and *PNAS*) and has calculated several measures to ascertain the impact of those papers, including the popular IF. All three journals tend to publish disproportionate number of Bioscience papers. When one takes into account the production of papers in different fields in the entire database, Bioscience papers are indeed overrepresented, as well as those from Astronomy and Geosciences. All other areas are underrepresented, with Professional fields, Humanities and Engineering being the most under-represented of all the areas. Interestingly, even though severely underrepresented, papers from Professional fields and Engineering are among the highest-impact areas in *PNAS* and their measure of prestige (how many times the IF is higher than that of the entire field) rivals that of articles from some areas published in *Nature* and *Science*. Also, while Astronomy, has come up at the bottom when it comes to the prestige because Astronomy papers from the entire field already have a relatively high IF, the three general science journals are highest ranked in terms of the impact factor. It is also interesting that Chemistry papers are only a small percentage of the papers in all three journals, but they have the highest IF in both *Science* and *Nature*. Also, Medicine papers are attracting citations above the overall IF for their respective journals. Bioscience, on the other hand, which is the most prevalent in all three journals, has IFs that is the closest to the IFs of the whole journals. Does it mean that editors of these journals are a little bit less selective when it comes to Bioscience papers, so papers from this area are doing comparable to their citation performance in the disciplinary venues? On the other hand, in the areas where the editors are more selective, and where there is probably much higher competition to get the publication accepted, published papers are doing significantly better, not only compared to the papers in their respective fields, but compared to papers from other fields published in these top venues.

To conclude, the answer to the question by which we motivated this study: do different areas have different impacts in these journals, is a resounding yes, with the ratio between highest and lowest IFs being 2.5 times in *Nature* and *Science* and ~6 times in *PNAS*.

## ACKNOWLEDGMENT:
This work uses Web of Science data by Clarivate Analytics provided by the Indiana University Network Science Institute and the Cyberinfrastructure for Network Science Center at Indiana University. This material is based upon work supported by the Air Force Office of Scientific Research under award number FA9550-19-1-0391. I dedicate this paper to the memory of my dear friend and colleague Judit Bar-Ilan. I think she would have been interested in this topic and I wish we had an opportunity to work on it together.

## REFERENCES
Ackerson, L. G., & Chapman, K. (2003). Identifying the role of multidisciplinary journals in scientific research. *College & Research Libraries, 64*(6), 468-478.
Archambault, É., & Larivière, V. (2009). History of the journal impact factor: Contingencies and consequences. *Scientometrics, 79*(3), 635-649.
Bar-Ilan, J. (2010). Rankings of Information and Library Science journals by JIF and by h-type indices. *Journal of Informetrics, 4*(2), 141-147.
Bar-Ilan, J. (2012). Journal report card. *Scientometrics, 92*(2), 249-260.
Bornmann, L., Werner, M., Gasparyan, A. Y., & Kitas, G. D. (2012). Diversity, value and limitations of the journal impact factor and alternative metrics. *Rheumatology International, 32*(7), 1861-1867.
Calcagno, V., Demoinet, E., Gollner, K., Guidi, L., Ruths, D., & de Mazancourt, C. (2012). Flows of research manuscripts among scientific journals reveal hidden submission patterns. *Science, 338*(6110), 1065-1069.
Csiszar, A. (2018). *The scientific journal: Authorship and the politics of knowledge in the nineteenth century*. Chicago: The University of Chicago Press.

De Rijcke, S., Wouters, P., Rushforth, A., Franssen, T., & Hammarfelt, B. (2016). Evaluation practices and effects of indicator use - a literature review. *Research Evaluation, 25*(2), 161-169.
Ding, J., Ahlgren, P., Yang, L., & Yue, T. (2018). Disciplinary structures in Nature, Science and PNAS: journal and country levels. *Scientometrics, 116*(3), 1817-1852.
DORA. (2012). San Francisco declaration of research assessment. Retrieved from <http://www.ascb.org/files/SFDeclarationFINAL.pdf>.
Fang, H. (2015). Classifying research articles in multidisciplinary science journals into subject categories. *Knowledge Organization, 42*(3), 139-153.
Fortunato, S., Bergstrom, C. T., Börner, K., Evans, J. A., Helbing, D., Milojević, S., . . . Barabási, A.-L. (2018). Science of science. *Science, 359*(6379), eaao0185.
Garfield, E. (1972). Citation analysis as a tool in journal evaluation. *Science, 178*, 471-479.
Garfield, E. (2006). The history and meaning of the journal impact factor. *JAMA, 295*(1), 90-93.
Glänzel, W., & Schubert, A. (2003). A new classification scheme of science fields and subfields designed for scientometric evaluation purposes. *Scientometrics, 56*(3), 357-367.
Glänzel, W., Schubert, A., & Czerwon, H. J. (1999). An item-by-item subject classification of papers published in multidisciplinary and general journals using reference analysis. *Scientometrics, 44*(3), 427-439.
Glänzel, W., Schubert, A., Schoepflin, U., & Czerwon, H. J. (1999). An item-by-item subject classification of papers published in journals covered by the SSCI database using reference analysis. *Scientometrics, 46*(3), 431-441.
Gordon, M. D. (1984). How authors select journals: A test of the reward maximization model of submission behavior. *Social Studies of Science, 14*, 27-43.
Haustein, S. (2012). *Multidimensional journal evaluation: Analyzing scientific periodicals beyond the impact factor*. Berlin: De Gruyter Saur.
Hicks, D. M., & Katz, J. S. (1996). Where is science going? *Science, Technology & Human Values, 21*(4), 379-406.
Ioannidis, J. P. (2006). Concentration of the most-cited papers in the scientific literature: analysis of journal ecosystems. *PloS one, 1*(1), e5.
Javitz, H., Grimes, T., Hill, D., Rapoport, A., Bell, R., Fecso, R., & Lehming, R. (2010). U.S. Academic Scientific Publishing. Working paper SRS 11-201. Arlington, VA: National Science Foundation, Division of Science Resources Statistics.
Johns, A. (1998). *The nature of the book: Print and knowledge in the making*. Chicago: The University of Chicago Press.
Kaneiwa, K., Adachi, J., Aoki, M., Masuda, T., Midorikawa, A., Tanimura, A., & Yamazaki, S. (1988). A comparison between the journals Nature and Science. *Scientometrics, 13*(3-4), 125-133.
Katz, J. S., & Hicks, D. (1995). *The classification of interdisciplinary journals: A new approach*. Paper presented at the Proceedings of the fifth international conference of the international society for scientometrics and informetrics, Rosary College, River Forest, IL.
Larivière, V., Kiermer, V., MacCallum, C., McNutt, M., Patterson, M., Pulverer, B., . . . Curry, S. (2016). A simple proposal for the publication of journal citation distributions. *Biorxiv*, doi:10.1101/062109.
López-Illescas, C., Noyons, E. C., Visser, M. S., De Moya-Anegón, F., & Moed, H. F. (2009). Expansion of scientific journal categories using reference analysis: How can it be done and does it make a difference? *Scientometrics, 79*(3), 473-490.
Meadows, A. J. (1998). *Communicating research*. San Diego: Academic Press.
Milojević, S. (revision under review). Practical method to reclassify Web of Science articles into unique subject categories and broad disciplines. *Quantitative Science Studies*.
Milojević, S., Radicchi, F., & Bar-Ilan, J. (2017). Citation success index - An intuitive pair-wise journal comparison metric. *Journal of Informetrics, 11*(1), 223-231.

Palmer, C. L. (1996). Information work at the boundaries of science: Linking library services to research practices. *Library Trends, 45*(2), 165-191.

Price, D. J. d. S. (1961). *Science since Babylon*. New Haven: Yale University Press.

Price, D. J. d. S. (1963). *Little science, big science*. New York: Columbia University Press.

Price, D. J. d. S. (1974). Society's needs in scientific and technical information. *Ciência da Informação, 3*(2), 97-103.

Radicchi, F., Fortunato, S., & Castellano, C. (2008). Universality of citation distributions: Toward an objective measure of scientific impact. *PNAS, 105*, 17268-17272.

Ravetz, J. R. (1971). *Scientific knowledge and its social problems*. New York: Oxford University Press.

Reedijk, J., & Moed, H. F. (2008). Is the impact of journal impact factors decreasing? *Journal of Documentation, 64*(2), 183-192.

Rousseau, R. (2002). Journal evaluation: Technical and practical issues. *Library Trends, 50*(3), 418-439.

Rousseau, S., & Rousseau, R. (2012). Interactions between journal attributes and authors' willingness to wait for editorial decisions. *Journal of the American Society for Information Science and Technology, 63*(6), 1213-1225.

Rushforth, A., & De Rijcke, S. (2015). Accounting for impact? The Journal Impact Factor and the making of biomedical research in the Netherlands. *Minerva, 53*(2), 117-139.

Seglen, P. O. (1992). The skewness of science. *Journal of the American Society for Information Science, 43*(9), 628-638.

Seglen, P. O. (1997). Why the impact factor of journals should not be used for evaluating research. *BMJ: British Medical Journal, 314*(7079), 498-502.

Sekara, V., Deville, P., Ahnert, S. E., Barabási, A. L., Sinatra, R., & Lehmann, S. (2018). The chaperone effect in scientific publishing. *Proceedings of the National Academy of Sciences, 115*(50), 12603-12607.

Stringer, M., Sales-Pardo, M., & Amaral, L. A. (2008). Effectiveness of journal ranking schemes as a tool for locating information. *PloS one, 3*(2), e1683.

Thelwall, M. (2012). Journal impact evaluation: A webometric perspective. *Scientometrics, 92*(2), 429-441.

Verma, I. M. (2015). Impact, not impact factor. *Proceedings of the National Academy of Sciences (PNAS), 112*(26), 7875-7876.

Waltman, L., & van Eck, N. J. (2012). A new methodology for constructing a publication-level classification system of science. *Journal of the American Society for Information Science and Technology, 63*(12), 2378-2392.

Ziman, J. (1969). Information, communication, knowledge. *Nature, 224*, 318-324.

Zuckerman, H., & Merton, R. K. (1971). Patterns of evaluation in science: Institutionalisation, structure and functions of the referee system. *Minerva, 9*(1), 66-100.